\documentclass[prl,twocolumn,superscriptaddress,showpacs,preprintnumbers]{revtex4}
\usepackage{amssymb}
\usepackage{graphicx}
\usepackage{dcolumn}
\usepackage{bm}
\usepackage{color}

\begin{document}

\title{ Competition between Anderson localization and antiferromagnetism in
correlated lattice fermion systems with disorder}
\author{Krzysztof Byczuk}
\affiliation{Theoretical Physics III, Center for Electronic Correlations and Magnetism,
Institute for Physics, University of Augsburg, D-86135 Augsburg, Germany}
\affiliation{Institute of Theoretical Physics, University of Warsaw, ul. Ho\.za 69,
PL-00-681 Warszawa, Poland}
\author{Walter Hofstetter}
\affiliation{Institut f\"ur Theoretische Physik, Johann Wolfgang Goethe-Universit\"at,
60438 Frankfurt/Main, Germany}
\author{Dieter Vollhardt}
\affiliation{Theoretical Physics III, Center for Electronic Correlations and Magnetism,
Institute for Physics, University of Augsburg, D-86135 Augsburg, Germany}
\date{\today }

\begin{abstract}
The magnetic ground state phase diagram of the disordered Hubbard
model at half-filling is computed in dynamical mean-field theory
supplemented with the  spin resolved, typical local density of
states. The competition between many-body correlations and disorder
is found to stabilize paramagnetic and antiferromagnetic metallic
phases at weak interactions. Strong disorder leads to Anderson
localization of the electrons and suppresses the antiferromagnetic
long-range order. Slater and Heisenberg antiferromagnets respond
characteristically different to disorder. The results can be tested
with cold fermionic atoms loaded into optical lattices.
\end{abstract}

\maketitle

Interacting quantum many-particle systems with disorder pose fundamental
challenges for theory and experiment not only in condensed matter physics
\cite{Mott90,Lee85,Altshuler85,Belitz94,Abrahams01}, but most recently also
in the field of cold atoms in optical lattices \cite%
{Lewenstein07,Fallani07,Billy08,Roati08,White08}. Indeed, ultracold
gases have quickly developed into a fascinating new laboratory for
quantum many-body physics; see e.g.
\cite{Bloch08,Jaksch02,Greiner02,HTC,Koehl05,bose-fermi,Mott}. A
major advantage of cold atoms in optical lattices is the high degree
of controllability of the interaction and the disorder strength,
thereby allowing a detailed verification of theoretical predictions.

In particular, these quantum many-body systems will allow for the
first experimental investigation of the simultaneous presence of
strong interactions and strong disorder. This very interesting
parameter regime is not easily accessible in correlated electron
materials. Namely, at or close to half-filling where interaction
effects become particularly pronounced, strong disorder implies
fluctuations (e.g., of local energies) of the order of the
band-width, which usually leads to structural instabilities. These
limitations are absent in the case of cold atoms in optical lattices
where disorder can be tuned to become arbitrarily strong without
destroying the experimental setup. Since at half filling and in the
absence of frustration effects interacting fermions order
antiferromagnetically, several basic questions arise: (i) How is a
non-interacting, Anderson localized system at half filling affected
by a local interaction between the particles? (ii) How does an
antiferromagnetic insulator at half filling respond to disorder
which in the absence of interactions would lead to an Anderson
localized state? (iii) Do Slater and Heisenberg antiferromagnets
behave differently in the presence of disorder? In this Letter we
provide answers to the above questions by calculating the zero
temperature, magnetic phase diagram of the disordered
Hubbard model at half filling using Dynamical Mean-Field Theory (DMFT) \cite%
{metzner89,georges96,PT} with a geometric average over the disorder \cite%
{Dobrosavljevic97,Dobrosavljevic03,Schubert03,Byczuk05,byczuk05a} and
allowing for a spin-dependence of the density of states (DOS).

Antiferromagnetic (AF) long-range order is a generic property of interacting
lattice fermions with particle-hole symmetry, as exemplified by the Hubbard
model at half filling with nearest-neighbor hopping on a bipartite lattice
\cite{Penn66Langer69,Jarrell92,Pruschke05}. Such an instability is also
highly relevant for current and future experiments in optical lattices \cite%
{AF-atoms}, where the magnetic super-exchange energy scale has recently been
observed in a two-component bosonic mixture \cite{Trotzky08}. The influence
of disorder, e.g., due to fluctuating local potentials, on interacting
quantum particles is subtle and leads to a remarkably rich phase diagram
which was studied by a variety of numerical techniques \cite%
{Logan93,Ulmke95,Denteneer98,Singh98,Heidarian04,Andersen07,Pezzoli08}.
While previous investigations yielded important insights into the properties
of disordered Hubbard antiferromagnets in various regions of parameter
space, a comprehensive study, where effects due to Anderson localization,
genuine many-body correlations and AF order are treated within the same
non-perturbative theoretical framework, did not yet exist. To this end we
here employ the DMFT --- a non-perturbative approach to correlated lattice
fermions which accounts for the Mott-Hubbard metal-insulator transition
(MIT) and magnetic ordering --- in combination with a disorder average which
is able to detect Anderson localization on the one-particle level \cite%
{Dobrosavljevic97,Dobrosavljevic03,Schubert03}. Namely, by employing the
geometric rather than the arithmetic average over the disorder it is
possible to determine the \emph{typical} local DOS \cite{Anderson58} as a
dynamical mean field within the DMFT. This approach was recently employed to
calculate the paramagnetic phase diagram of the disordered Hubbard model
\cite{Byczuk05} and Falicov-Kimball model \cite{byczuk05a}. Thereby it was
possible to determine the MIT due to disorder (Anderson localization) and
interactions (Mott-Hubbard transition), respectively, as well as the
transition scenario caused by their simultaneous presence, within a unified
framework.

In the absence of frustration effects the Mott-Hubbard MIT is completely
hidden by AF long-range order \cite{georges96,Pruschke05}. To capture this
feature it is necessary to generalize the investigation and include AF
solutions of the Hubbard model with local disorder (Anderson-Hubbard model),
whose Hamiltonian is given by
\begin{equation}
H_{AH}= - \sum_{ ij \sigma }t_{ij} a_{i\sigma }^{\dagger }a_{j\sigma
}+\sum_{i\sigma }\epsilon _{i}n_{i\sigma }+U\sum_{i}(n_{i\uparrow}-\frac{1}{2%
}) (n_{i\downarrow }-\frac{1}{2}).  \label{1}
\end{equation}
Here $t_{ij}$ is the amplitude for hopping between the sites $i$ and $j$, $U$
is the on-site repulsion, $n_{i\sigma }=a_{i\sigma }^{\dagger }a_{i\sigma }^{%
{\phantom{\dagger}}}$ is the local fermion number operator with $a_{i\sigma
} $ ($a_{i\sigma }^{\dagger}$) as the annihilation (creation) operator of a
fermion with spin $\sigma$, and $\epsilon _{i}$ are random on-site energies.
In the following we work with a continuous probability distribution function
for $\epsilon _{i}$, i.e., $\mathcal{P}(\epsilon _{i})=\Theta (\Delta
/2-|\epsilon _{i}|)/\Delta ,$ with $\Theta $ as the step function. The
parameter $\Delta $ is a measure of the disorder strength. We consider a
bipartite lattice with equal number of fermions and lattice sites
(half-filled case). In the absence of disorder the Hamiltonian is then
explicitly particle-hole symmetric.

The Anderson-Hubbard model (\ref{1}) is solved within DMFT by mapping it
onto single-impurity Anderson Hamiltonians with different $\epsilon_{i}$
\cite{georges96,Ulmke95}. For each random on-site energy $\epsilon _{i}$,
where $i$ belongs to one of the sublattices $s=$A or B, we calculate the
local Green function $G_{\sigma s}(\omega ,\epsilon _{i})$. From this
quantity we obtain the geometrically averaged local DOS $\rho _{\sigma s}^{%
\mathrm{geom}}(\omega )=\exp \left[ \langle \ln \rho _{\sigma s}(\omega ,
\epsilon_i)\rangle \right] $ \cite{Dobrosavljevic97,Dobrosavljevic03}, where
$\rho _{\sigma s}(\omega , \epsilon_i)=-\mathrm{Im}G_{\sigma s}(\omega
,\epsilon _{i})/\pi $, and $\langle O \rangle =\int d\epsilon _{i}\mathcal{P}%
(\epsilon _{i})O(\epsilon _{i})$ denotes the arithmetic average of $%
O(\epsilon_i)$. For comparison, the arithmetically averaged local DOS $%
\rho_{\sigma s}^{\mathrm{arith}}(\omega)= \langle \rho _{\sigma s
}(\omega , \epsilon_i)\rangle$ is also computed. The averaged local
Green function is then obtained from the Hilbert transform
$G_{\sigma s}^{\alpha}(\omega )=\int d\omega ^{\prime }\rho_{\sigma
s} ^{\alpha}(\omega ^{\prime})/(\omega -\omega ^{\prime })$, where
$\alpha = \mathrm{geom}$ (arith) denotes the geometric (arithmetic)
average. The local self-energy $\Sigma _{\sigma s}^{\alpha} (\omega
)$ is determined from the $\mathbf{k}$-integrated Dyson equation
$\Sigma_{\sigma s}^{\alpha}(\omega )=\omega -\eta_{\sigma
a}^{\alpha} (\omega )-1/G_{\sigma s}^{\alpha}(\omega )$ where $\eta
^{\alpha}_{\sigma s} (\omega)$ is the hybridization function of the
effective Anderson Hamiltonian. The latter quantity provides the
position and the resonant broadening of single-site quantum levels
and may be interpreted as a molecular mean-field which describes the
effect of all other sites within the DMFT. The self-consistent DMFT
equations are closed by the Hilbert transform of the Green function
on a bipartite lattice
\begin{equation}
G_{\sigma s}^{\alpha}(\omega )=\int d\epsilon \; \frac{N_{0}(\epsilon )}{%
\left[\omega -\Sigma_{\sigma s}^{\alpha} (\omega )-\frac{\epsilon^2}{%
\omega-\Sigma _{\sigma \bar{s}}^{\alpha}(\omega)}\right]},
\end{equation}
where $N_{0}(\epsilon )$ is the non-interacting DOS and $\bar{s}$ denotes
the sublattice opposite to $s$. In the following we choose a model DOS, $%
N_{0}(\epsilon )=2\sqrt{D^{2}-\epsilon ^{2}}/\pi D^2$, with bandwidth $W=2D$%
, and set $W=1$. For this DOS and a bipartite lattice the local
Green function and the hybridization function are connected by the
simple algebraic
relation $\eta_{\sigma s} ^{\alpha}(\omega )=D^{2}G_{\sigma \bar{s}%
}^{\alpha}(\omega )/4$ \cite{georges96}. The DMFT equations are solved at
zero temperature by the numerical renormalization group technique \cite%
{NRG}, which allows us to calculate the geometric
or arithmetic average of the local DOS in each iteration loop.

\begin{figure}[tbp]
\includegraphics [clip,width=8cm,angle=-00]{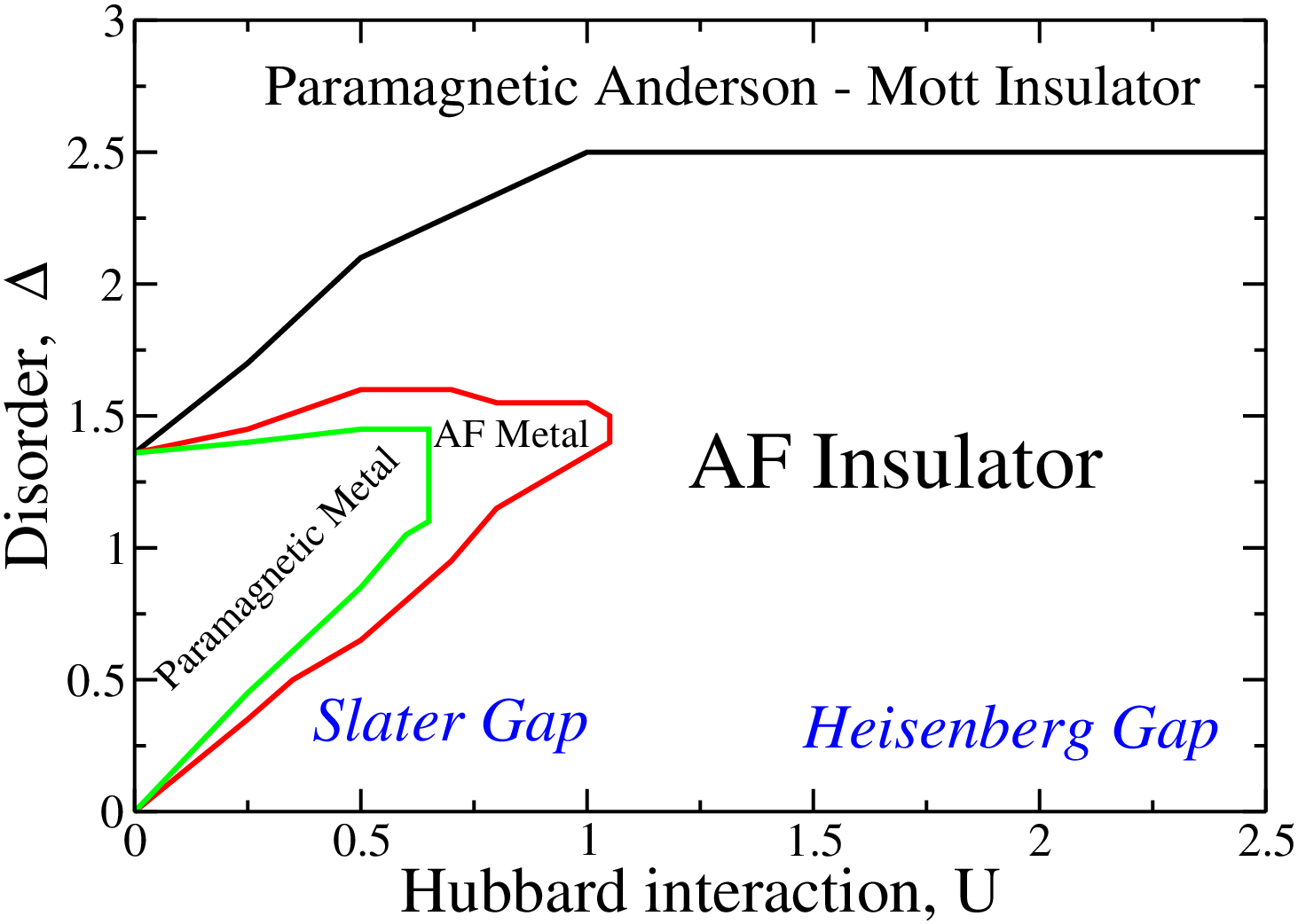}
\caption{Magnetic ground state phase diagram of the Anderson-Hubbard
model at half-filling as calculated by DMFT with a spin resolved
local DOS (see text).} \label{fig1}
\end{figure}

To characterize the ground state of the Hamiltonian (\ref{1}) the following
quantities are computed: The local DOS $\rho_{\sigma s}^{\alpha}(\omega )$
for a given sublattice $s$ and spin direction $\sigma$, the total DOS for a
given sublattice $s$ at the Fermi level $N_s^{\alpha}(0)\equiv
\sum_{\sigma}\rho_{\sigma s}^{\alpha}(\omega=0)$, and the staggered
magnetization $m_{\mathrm{AF}}^{\alpha}=|n_{\uparrow A}^{\alpha}-n_{\uparrow
B}^{\alpha}|$, where $n_{\sigma s}^{\alpha}=\int_{-\infty}^0 d\omega \rho
_{\sigma s}^{\alpha}(\omega )$ is the on-site particle density on each
sublattice. The possible phases of the Anderson-Hubbard model can then be
classified as follows: The systems is a \newline
i) paramagnetic metal if $N_s^{\mathrm{geom}}(0)\neq 0$ and $m_{\mathrm{AF}%
}^{\mathrm{geom}}=0$, \newline
ii) AF metal if $N_s^{\mathrm{geom}}(0)\neq 0$ and $m_{\mathrm{AF}}^{\mathrm{%
geom}}\neq 0$, \newline
iii) AF insulator if $N_s^{\mathrm{geom}}(0)=0$ and $m_{\mathrm{AF}}^{%
\mathrm{geom}}\neq 0$ but $N_{ s}^{\mathrm{geom}}(\omega )\neq 0$ for some $%
\omega \neq 0$, and \newline
iv) paramagnetic Anderson-Mott insulator if $N_{s}^{\mathrm{geom}}(\omega
)=0 $ for all $\omega$.

\begin{figure}[tbp]
\includegraphics [clip,width=8cm,angle=-00]{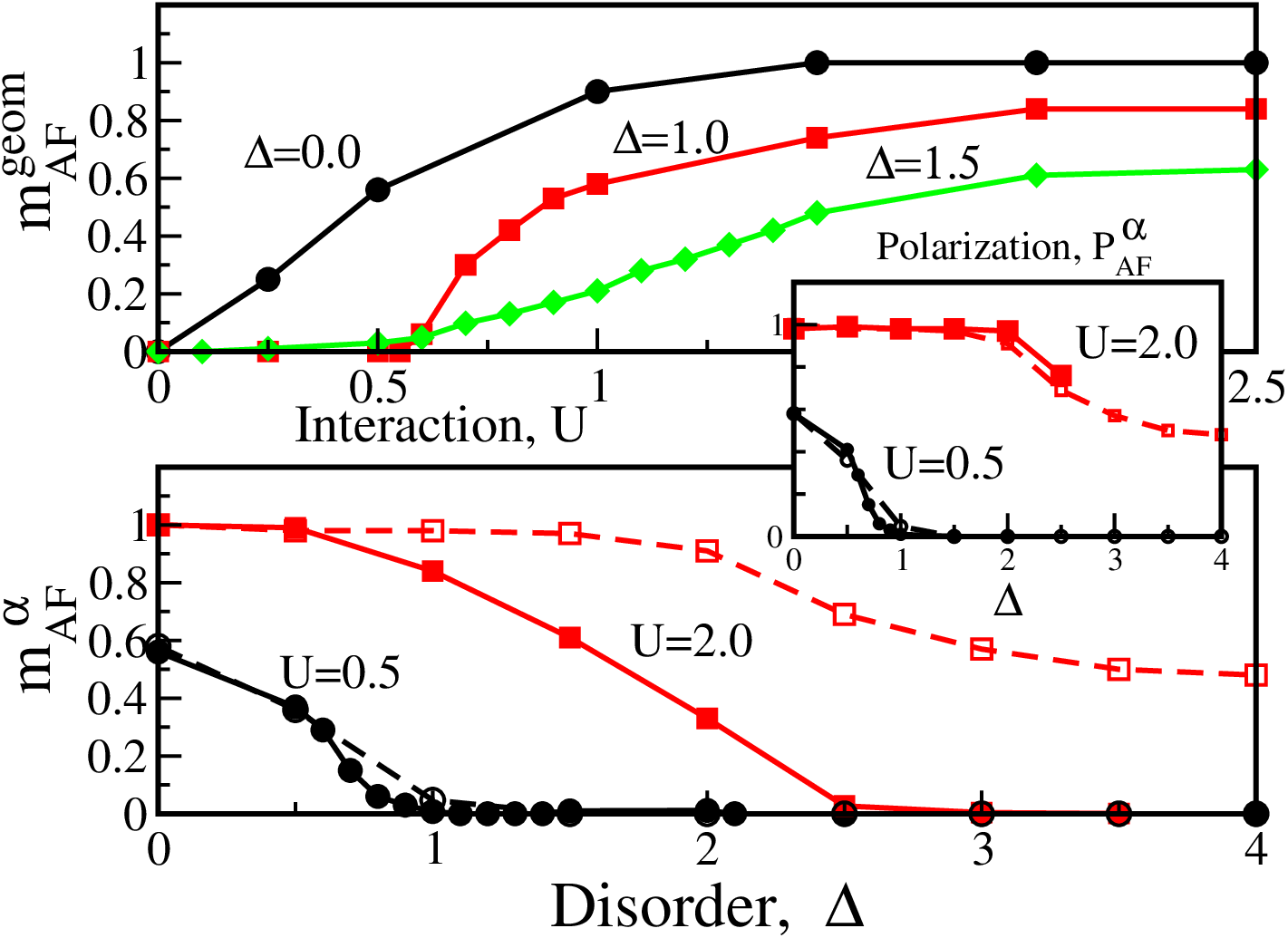}
\caption{Upper panel: Staggered magnetization
$m_{\mathrm{AF}}^{\mathrm{geom}}$ as a
function of interaction $U$. Lower panel: $m_{\mathrm{AF}}^{\protect%
\alpha }$, $\protect\alpha =\mathrm{geom}/\mathrm{arith}$, as a
function of disorder $\Delta $. Inset: Polarization
$P_{\mathrm{AF}}^{\protect\alpha }$ as a function of disorder.
Dashed lines present results obtained by arithmetic averaging. }
\label{fig3}
\end{figure}

The ground state phase diagram of the Anderson-Hubbard model (\ref{1})
obtained by this classification is shown in Fig.~\ref{fig1}. Depending on
whether the interaction $U$ is weak or strong
the response of the system to disorder is found to be very different.
In particular, at strong interactions,
$U/W\gtrsim 1$,
there exist only two phases, an AF insulating phase at weak disorder, $%
\Delta /W\lesssim 2.5$, and a paramagnetic Anderson-Mott insulator at strong
disorder, $\Delta /W\gtrsim 2.5$. The transition between these two phases is
continuous. Namely, the local DOS and the staggered magnetization both
decrease gradually as the disorder $\Delta $ increases and vanish at their
mutual boundary (lower panel of Fig.~\ref{fig3}). By contrast, the phase
diagram for weak interactions, $U/W\lesssim 1,$ has a much richer structure
(Fig.~\ref{fig1}). In particular, for weak disorder a \textit{paramagnetic}
metallic phase is stable. It is separated from the AF insulating phase at
large $U$ by a narrow region of \textit{AF} \textit{metallic} phase.

To better understand the nature of the AF phases in the phase
diagram we take a look at the staggered magnetization
$m_{\mathrm{AF}}^{\alpha }$.
The dependence of $m_{\mathrm{AF}%
}^{\mathrm{geom}}$ on $U$ is shown in the upper panel of
Fig.~\ref{fig3} for several values of the disorder $\Delta $. In
contrast to the non-disordered case a finite interaction strength
$U>U_{c}(\Delta )$ is needed to stabilize the AF long-range order
when disorder is present. The staggered magnetization saturates at
large $U$ for both averages; the maximal values depend on the
disorder strength. In the lower panel of Fig.~\ref{fig3} the
dependence of $m_{\mathrm{AF}}^{\alpha }$ on the disorder $\Delta $
is shown for different interactions $U$. Only for small $U$ do the
two averages yield approximately the same results.
Another useful quantity is the polarization  $P_{\mathrm{AF}}^{\alpha }=m_{%
\mathrm{AF}}^{\alpha }/I^{\alpha }$, where $I^{\alpha
}=\int_{-\infty }^{+\infty }\sum_{\sigma s}\rho _{\sigma s}^{\alpha
}(\omega )d\omega /2$ is the total spectral weight of $\rho _{\sigma
s}^{\alpha }(\omega )$. It allows one to investigate the
contribution of the point-like spectrum of the Anderson localized
states to the magnetization. This provides important information
about the spectrum since with increasing disorder more and more
one-particle states of the many-body system are transferred from the
continuous to the point-like spectrum. For weak interactions
($U=0.5$) the decrease of the polarization with increasing disorder
$\Delta $ obtained with geometric or arithmetic averaging is the same (see inset in Fig.~%
\ref{fig3}). Since arithmetic averaging does not treat states from
the point-like spectrum correctly, the decrease of
$m_{\mathrm{AF}}^{\alpha }$ (which is also the same for the two
averages, see lower panel of Fig.~\ref{fig3}) must be attributed to
disorder effects involving only the continuous spectrum. At larger
$U$ the polarization is constant up to the transition from the AF
insulator to the paramagnetic Anderson-Mott insulator. In the latter
phase the polarization is undefined, because the continuous spectrum
does not contribute to $I_{\mathrm{AF}}^{\mathrm{geom}}$.

\begin{figure}[tbp]
\includegraphics [clip,width=7.5cm,angle=-00]{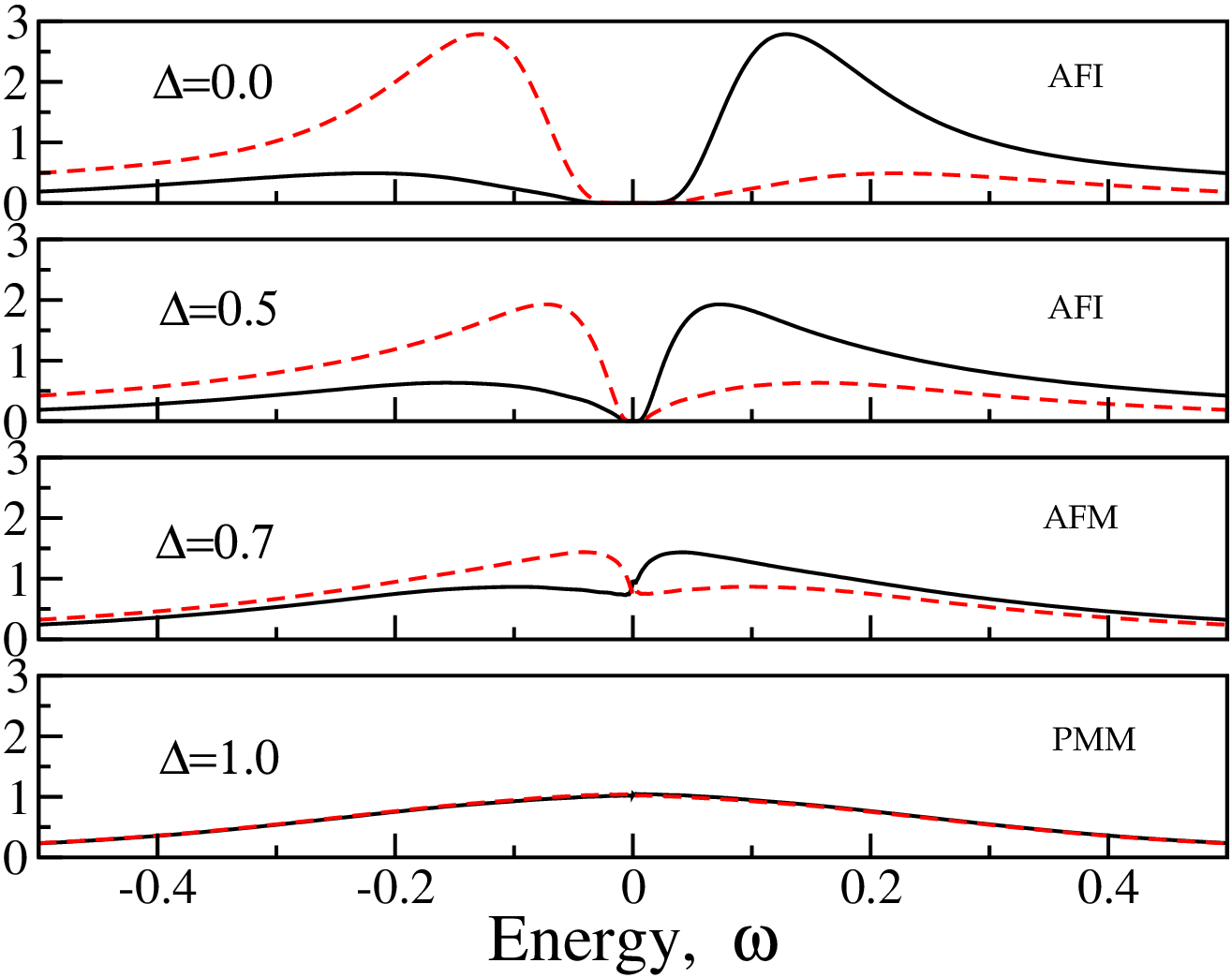} %
\includegraphics [clip,width=7.5cm,angle=-00]{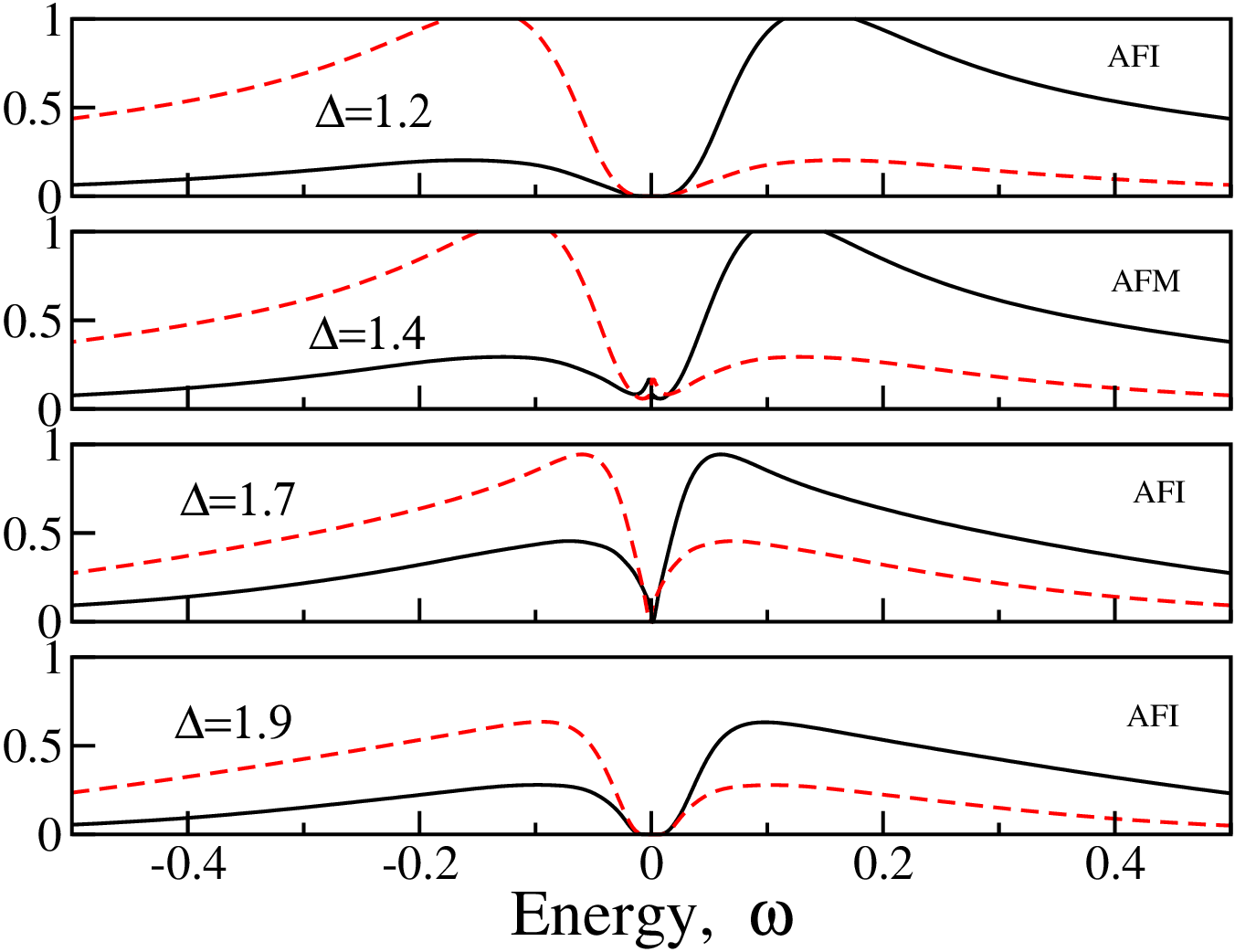}
\caption{Typical local DOS
as a function of disorder $\Delta$ for interaction $%
U=0.5$ (upper panel) and $U=1.0$ (lower panel). Solid and dashed
lines represent opposite spin directions. } \label{fig6}
\end{figure}

The AF metallic phase is long-range ordered, but there is no gap since the
disorder leads to a redistribution of spectral weight. In Fig.~\ref{fig6}
the local DOS in the vicinity of the transitions between the paramagnetic
metal, the AF insulator and the AF metal at $U=0.5$ (upper panel), and the
transitions between the AF insulator, the AF metal and back into the AF
insulator at $U=1.0$ (lower panel) are shown. The paramagnetic metal, where $%
\rho _{\sigma s}^{\alpha }(\omega )=\rho _{\sigma \bar{s}}^{\alpha }(\omega )
$, is seen to be stable only for weak interactions.

In the absence of disorder the AF insulating phase has a small ("Slater")
gap at $U/W<1$ and a large ("Heisenberg") gap at $U/W>1$. These limits can
be described by perturbation expansions in $U$ and $1/U$ around the symmetry
broken state of the Hubbard and the corresponding Heisenberg model,
respectively. In agreement with earlier studies \cite{pruschke} our results
for $m_{\mathrm{AF}}$ (upper panel of Fig.~\ref{fig3}) show that there is no
sharp transition between these limits, even when disorder is present. This
may be attributed to the fact that both limits are described by the same
order parameter. However, the phase diagram (Fig.~\ref{fig1}) shows that the
two limits \emph{can} be distinguished by their overall response to
disorder. Namely, the reentrance of the AF metallic phase at $%
\Delta/W\gtrsim 1$ occurs only within the Slater AF insulating phase.

The magnetic structure of the Anderson-Mott insulator cannot be
determined by the method used here since it describes only the
continuous part of the spectra and not the point spectrum. However,
only the paramagnetic solution should be expected to be stable
because the kinetic exchange interaction responsible for the
formation of the AF metal is suppressed by the disorder. This does
not exclude the possibility of Griffiths phase-like AF domains
\cite{Griffiths}.

It is interesting to note that even the DMFT with an arithmetic average
finds a disordered AF metal \cite{Ulmke95,Singh98}. However, the
arithmetically averaged local DOS incorrectly predicts both the paramagnetic
metal and the AF metal to remain stable for arbitrarily strong disorder.
Only a computational method which is sensitive to Anderson localization,
such as the DMFT with geometrically averaged local DOS employed here, is
able to detect the suppression of the metallic phase for $\Delta /W\gtrsim
1.5$ and the appearance of the paramagnetic Anderson-Mott insulator at large
disorder $\Delta $ already on the one-particle level.

In conclusion, we computed the ground state phase diagram of the
Anderson-Hubbard model at half filling within a non-perturbative
approach which can treat interactions and disorder of arbitrary
strength and is sensitive to Anderson localization on the
one-particle level. For low disorder and weak interactions
paramagnetic and antiferromagnetic metallic phases become stable,
with a reentrant behavior of the latter phase. Slater and Heisenberg
antiferromagnets can be distinguished by their very different
response with respect to disorder. Experiments with cold fermionic
atoms loaded into optical lattices will be able to test these
predictions and check the accuracy of the theoretical approach
employed here.

We thank R.~Bulla for useful discussions. This work was supported in
part by the Sonderforschungsbereich 484 and the Forschergruppe FOR
801 of the Deutsche Forschungsgemeinschaft.

%%%%%%%%%%%%%%%%%%%%%%%%%%%%%%%%%%%%%%%%%%%%%%%

\end{document}